\documentclass[12pt]{article}

\usepackage{amsfonts}
\usepackage{amssymb}
\usepackage{amscd}

\def\D{\hbox{D\kern-.73em\raise.25ex\hbox{-}\raise-.25ex\hbox{ }}}
 \def\d{\hbox{d\kern-.33em\raise.75ex\hbox{-}\raise-.75ex\hbox{}}}

\def\GGG{\frak G }
\def\gr3{\GGG\,(\SSS_3)}

\def\gr2{\GGG\,(\SSS_2)}

\def\SSS{\frak S}

\def\ed{\end{document}}
\def\beq{\begin{equation}}
\def\eeq{\end{equation}}
\def\bea{\begin{eqnarray}}
\def\eea{\end{eqnarray}}
\def\ba{\begin{array}}
\def\ea{\end{array}}
\def\bi{\begin{itemize}}
\def\ei{\end{itemize}}

\def\noi{\noindent}
\def\nn{\nonumber}

\newcommand{\bp}{\noindent\begin{minipage}[c]}
\newcommand{\ep}{\end{minipage}}

\date{}

\begin{document}

\title{\Large\bf  $p$-Adic Modelling of the Genome and the Genetic Code}

\author{Branko
Dragovich$^a$\,\footnote{\textsf{\,
E-mail:\,dragovich@phy.bg.ac.yu}}\,\, and Alexandra
Dragovich$^b$\\ {} \\
 {\it $^a$Institute of Physics} \\ { \it P.O. Box 57,  11001
Belgrade, Serbia} \\ {\it $^b$Vavilov Institute of General
Genetics}\\ {\it Gubkin St. 3, \, 119991  Moscow, \,Russia}}

\maketitle

%\bigskip

%\shortauthors{B. Dragovich and A. Dragovich}

%\received{00 July 2007} %\revised{00 Month 2004}

\begin{abstract}

The present paper is devoted to foundations of $p$-adic modelling
in genomics. Considering nucleotides, codons, DNA and RNA
sequences, amino acids, and proteins as information systems, we
have formulated the corresponding $p$-adic formalisms for their
investigations. Each of these systems has its characteristic prime
number  used for construction of the related information space.
Relevance of this approach is illustrated by some examples. In
particular, it is shown that degeneration of the genetic code is a
$p$-adic phenomenon. We have also put forward a hypothesis on
evolution of the genetic code assuming that primitive code was
based on single nucleotides and chronologically first four amino
acids. This formalism of $p$-adic genomic information systems can
be implemented in computer programs and applied to various
concrete cases.
\end{abstract}

%\maketitle

\section{Introduction}

Living organisms seem to be the most complex, interesting and
significant objects regarding all substructures of the universe.
One of the essential characteristics that differ a living organism
from all other material systems is related to its genome. The
genome of an organism is its whole hereditary information encoded
in  the desoxyribonucleic acid (DNA), and contains  both  genes
and non-coding sequences. In some viruses, which are between
living and non-living objects, genetic material is encoded in the
ribonucleic acid (RNA). Investigation of the entire genome is the
subject of genomics. The human genome, which presents all genetic
information of the {\it Homo sapiens}, is composed of more than $3
\cdot 10^9$ DNA base pairs and contains more than  $3 \cdot 10^4$
genes \cite{watson}. For a more detailed and comprehensive
information on molecular biology aspects of DNA, RNA and the
genetic code one can use Ref. \cite{watson}. To have a
self-contained exposition we shall briefly review  some necessary
basic properties of genomics.

The DNA  is a macromolecule composed of two polynucleotide chains
with a double-helical structure. Nucleotides consist of a base, a
sugar and a phosphate group. Helical backbone is a result of the
sugar and phosphate groups. There are four bases and they are
building blocks of the genetic information. They are called
adenine (A), guanine (G), cytosine (C) and thymine (T). Adenine
and guanine are derived from purine, while cytosine and thymine
from pyrimidine. In the sense of information, the nucleotide and
its base present the same object. Nucleotides are arranged along
chains of double helix through base pairs A-T and C-G bonded by 2
and 3 hydrogen bonds, respectively. As a consequence of this
pairing there is an equal number of cytosine and guanine as well
as the equal rate of adenine and thymine. DNA is packaged in
chromosomes which are localized in the nucleus of the eukaryotic
cells.

The main role of  DNA is to storage genetic information and there
are two main processes to exploit this information. The first one
is replication, in which  DNA duplicates giving two new DNA
containing the same information as the original one. This is
possible owing to the fact that each of two chains contains
complementary bases of the other one. The second process is
related to the gene expression, i.e. the passage of DNA gene
information to proteins. It performs by the messenger ribonucleic
acid (mRNA), which is usually a single polynucleotide chain. The
mRNA is synthesized during the first part of this process, known
as transcription, when nucleotides C, A, T, G from DNA are
respectively transcribed into their complements G, U, A, C in
mRNA, where T is replaced by U (U is the uracil, which is a
pyrimidine). The next step in gene expression is translation, when
the information coded by codons in the mRNA  is translated into
proteins. In this process also participate  transfer tRNA and
ribosomal rRNA.

Codons are ordered  trinucleotides composed of C, A, U (T) and G.
Each of them presents an information which controls use of one of
the 20 standard amino acids or stop signal in synthesis of
proteins.

Protein synthesis in all eukaryotic cells performs in the
ribosomes of the cytoplasm. Proteins \cite{finkelshtein} are
organic macromolecules composed of amino acids arranged in a
linear chain. Amino acids are molecules that consist of  amino,
carboxyl and R (side chain) groups. Depending on R group there are
20 standard amino acids. These amino acids are joined together by
a peptide bond. Proteins are substantial ingredients of all living
organisms participating in various processes in cells and
determining the phenotype of an organism. In the human body there
may be about 2 million different proteins. The study of proteins,
especially their structure and functions, is called proteomics.
The proteome is the entire set of proteins in an organism.

The sequence of amino acids in a protein is determined by sequence
of codons contained in DNA genes.  The relation between codons and
amino acids is known as the {\it genetic code}. Although there are
at least 16 codes (see, e.g. \cite{sorba1}), the most important
are two of them: the standard (eukaryotic) code and the vertebral
mitochondrial code.

In the sequel we shall mainly have in mind the vertebral
mitochondrial code, because it is a simple one and the others may
be regarded as its slight modifications. It is obvious that there
are $4 \times 4\times 4 = 64$  codons. However (in the vertebral
mitochondrial code), $60$ of them are distributed on the $20$
different amino acids and $4$ make stop codons, which serve as
termination signals. According to experimental observations, two
amino acids are coded by six codons, six amino acids by four
codons, and twelve amino acids by two codons. This property that
some amino acids are coded by more than one codon is known as {\it
genetic code degeneracy}. This degeneracy is a very important
property of the genetic code and gives an efficient way to
minimize errors  caused by mutations.

Since there is in principle a huge number (between $10^{71}$ and
$10^{84}$  \cite{hornos}) of all possible assignments between
codons and amino acids,  and only a very small number  of them is
represented in living cells, it has been a persistent theoretical
challenge to find an appropriate model explaining contemporary
genetic codes. An interesting model based on the quantum algebra
$\mathcal{U}_q (sl(2)\oplus sl(2))$ in the $q \to 0$ limit was
proposed as a symmetry algebra for the genetic code (see
\cite{sorba1} and references therein). In a sense this approach
mimics quark model of baryons. To describe correspondence between
codons  and amino-acids, it was constructed an operator which acts
on the space of codons and its eigenvalues are related to amino
acids. Besides some successes of this approach, there is a problem
with rather many parameters in the operator. There are also papers
(see, e.g. \cite{hornos}, \cite{forger} and \cite{bashford})
starting with 64-dimensional irreducible representation of a Lie
(super)algebra and trying to connect multiplicity of codons with
irreducible representations of subalgebras arising in a chain of
symmetry breaking. Although interesting as an attempt to describe
evolution of the genetic code  these Lie algebra approaches did
not succeed to get its modern form.   For a very brief review of
these and some other theoretical approaches to the genetic code
one can see Ref. \cite{sorba1}. There is still no generally
accepted explanation of the genetic code.

It is worth recalling emergence of special theory of relativity
and quantum mechanics. They both appeared as a result of
unsatisfactory attempts to extend classical concepts to new
physical phenomena, and introduction of new physical ideas and
mathematical methods. Although far from everyday experience these
new two theories describe physical reality quite successfully. We
believe that similar situation should happen in theoretical
description of living processes in biological organisms.
Ultrametric and $p$-adic methods seem to be very promising tools
in further investigation of the life.

Modelling of the genome, the genetic code and proteins  is a
challenge as well as an opportunity for applications of $p$-adic
mathematical physics. Recently \cite{dragovich1} we introduced a
$p$-adic approach to  DNA and RNA sequences, and to the genetic
code. The central point of our approach is an appropriate
identification of four nucleotides with  digits $1,\, 2, \, 3, \,
4$ of $5$-adic integer expansions and application of $p$-adic
distances between obtained numbers. $5$-Adic numbers with three
digits form $64$ integers which correspond to  $64$ codons. In
\cite{dragovich1a} we analyzed $p$-adic degeneracy of the genetic
code. As one of the main results that we have obtained is
explanation of the structure of the genetic code degeneracy using
$p$-adic distance between codons. A similar approach to the
genetic code was considered on diadic plane \cite{kozyrev}.

Let us mention that $p$-adic models in mathematical physics have
been actively considered since 1987 (see \cite{freund},
\cite{vladimirov1} for early reviews and \cite{dragovich2},
\cite{dragovich3} for some recent reviews). It is worth noting
that $p$-adic models with pseudodifferential operators have been
successfully applied to interbasin kinetics of proteins
\cite{avetisov3}.  Some $p$-adic aspects of cognitive,
psychological and social phenomena have been also considered
\cite{khrennikov1}. The recent  application of $p$-adic numbers in
physics and related branches of sciences is reflected in the
proceedings of the 2nd International Conference on $p$-Adic
Mathematical Physics \cite{proceedings}.

This paper is devoted to the further $p$-adic modelling  of the
genome as well as to $p$-adic roots of the  genetic code evolution
based on approach introduced in \cite{dragovich1} and considered
in \cite{dragovich1a}.

\bigskip

\section{$p$-Adic Genome}

In  Introduction we presented a brief review of the genome and the
genetic code, as well as  some motivations for their $p$-adic
theoretical investigations. To consider $p$-adic properties of the
genome and the genetic code in a self-contained way we shall also
recall some mathematical preliminaries.

\subsection{Some mathematical preliminaries and $p$-adic codon space}

As a new tool to study the Diophantine equations, $p$-adic numbers
are introduced by German mathematician Kurt Hensel in 1897. They
are involved in many branches of modern mathematics, either as
rapidly developing topics or as suitable applications. An
elementary introduction to $p$-adic numbers  can be found in the
book \cite{gouvea}. However, for our purposes we will use here
only a small portion of $p$-adics, mainly some finite sets of
integers and ultrametric distances between them.

Let us introduce the  set of natural numbers

\beq \mathcal{C}_5\, [64] = \{ n_0 + n_1\, 5 + n_2\, 5^2 \,:\,\,
n_i = 1, 2, 3, 4 \}\,,   \label{2.1}\eeq where $n_i$ are digits
related to nucleotides by the following assignments: C (cytosine)
= 1,\, A (adenine) = 2,\, T (thymine) = U (uracil) = 3,\, G
(guanine) = 4. This is a finite expansion to the base $ 5$. It is
obvious that $5$ is a prime number and that the set $\mathcal{C}_5
[64]$ contains $64$ numbers between $31$ and $124$ in the usual
base $10$. In the sequel we shall often denote elements of
$\mathcal{C}_5 [64]$ by their digits to the base $5$ in the
following way: $ n_0 + n_1\, 5 + n_2\, 5^2 \, \equiv n_0\, n_1\,
n_2$. Note that here ordering of digits is the same as in the
expansion, i.e this ordering is opposite to the usual one. There
is now evident one-to-one correspondence between codons in
three-letter  notation and number $n_0\, n_1\, n_2$
representation.

It is also often important to know a distance between numbers.
Distance can be defined by a norm. On the set $\mathbb{Z}$ of
integers  there are two kinds of nontrivial norm: usual absolute
value $|\cdot|_\infty$ and $p$-adic absolute value $|\cdot|_p$ ,
where $p$ is any prime number. The usual absolute value is well
known from elementary  mathematics and the corresponding distance
between two numbers $x$ and $y$ is $d_\infty (x, y) =
|x-y|_\infty$.

The $p$-adic absolute value is related to the divisibility of
integers by prime numbers. Difference of two integers is again an
integer. $p$-Adic distance between two integers can be understood
as a measure of  divisibility by $p$ of their difference (the more
divisible, the shorter). By definition, $p$-adic norm of an
integer  $m \in \mathbb{Z}$, is $|m|_p = p^{-k}$, where $ k \in
\mathbb{N} \bigcup \{ 0\}$ is degree of divisibility of $m$ by
prime $p$ (i.e. $m = p^k\, m'\,, \,\, p\nmid m'$) and $|0|_p =0.$
This norm is a mapping from $\mathbb{Z}$ into non-negative
rational numbers and has the following properties:

(i) $|x|_p \geq 0, \,\,\, |x|_p =0$ if and only if $x = 0$,

(ii) $|x\, y|_p = |x|_p \,  |y|_p \,,$

(iii) $|x + y|_p \leq \, \mbox{max}\, \{ |x|_p\,, |y|_p \} \leq
|x|_p + |y|_p $ for all $x \,, y \in \mathbb{Z}$.

\noindent Because of the strong triangle inequality $|x + y|_p
\leq \, \mbox{max} \{ |x|_p\,, |y|_p \}$, $p$-adic absolute value
belongs to non-Archimedean (ultrametric) norm. One can easily
conclude that $0 \leq |m|_p \leq 1$ for any $m\in \mathbb{Z}$ and
any prime $p$.

$p$-Adic distance between two integers $x$ and $y$ is
\begin{equation}
d_p (x\,, y) = |x - y|_p \,.    \label{2.2}
\end{equation}
Since $p$-adic absolute value is ultrametric, the $p$-adic distance
(\ref{2.2}) is also ultrametric, i.e. it satisfies
\begin{equation}
d_p (x\,, y) \leq\, \mbox{max}\, \{ d_p (x\,, z) \,, d_p (z\,, y) \}
\leq d_p (x\,, z) + d_p (z\,, y) \,, \label{2.3}
\end{equation}
where $x, \, y$ and $z$ are any three integers.

The above introduced set $\mathcal{C}_5\, [64]$ endowed by
$p$-adic distance we shall call {\it $p$-adic codon space}, i.e.
elements of $\mathcal{C}_5\, [64]$ are codons denoted by $n_0 n_1
n_2$. $5$-Adic distance between two codons $a, b \in \mathcal{C}_5
\, [64]$ is

\beq d_5 (a,\, b) = |a_0 + a_1 \, 5 + a_2 \, 5^2 - b_0 - b_1 \, 5
- b_2 \, 5^2 |_5 \,,   \label{2.4} \eeq where $a_i ,\, b_i \in \{
1, 2, 3, 4\}$. When $a \neq b$ then $d_5 (a,\, b)$ may have three
different values:
\begin{itemize}
\item $d_5 (a,\, b) = 1$ if $a_0 \neq b_0$, \item $d_5 (a,\, b) =
1/5$ if $a_0 = b_0 $ and $a_1 \neq b_1$, \item $d_5 (a,\, b) =
1/5^2$ if $a_0 = b_0 \,, \,\,a_1 = b_1$ and $a_2 \neq b_2 $.
\end{itemize}
 We
see that the largest $5$-adic distance between codons is $1$ and
it is maximum $p$-adic distance on $\mathbb{Z}$. The smallest
$5$-adic distance on the codon space is $5^{-2}$. Let us also note
that $5$-adic distance depends only on the first two nucleotides
of different codons.

If we apply real (standard) distance $d_\infty (a,\, b) = |a_0 +
a_1 \, 5 + a_2 \, 5^2 - b_0 - b_1 \, 5 - b_2 \, 5^2 |_\infty $,
then third nucleotides $a_2$ and $b_2$ would play more important
role than those at the second position (i.e $a_1 \, \mbox{and} \,
b_1$), and nucleotides $a_0$ and $b_0$ are of the smallest
importance. At real $\mathcal{C}_5 [64]$ space distances are also
discrete, but take values $1,\, 2,\, \cdots , 93$. The smallest
real and the largest $5$-adic distance are equal $1$. While real
distance describes metric of the ordinary physical space, this
$p$-adic one serves to describe ultrametricity  of the codon
space.

It is worth emphasizing that the metric role of digits depends on
their position in number expansion and it is quite opposite  in
real and $p$-adic cases.  We shall see later, when we consider the
genetic code, that the first two nucleotides in a codon are more
important than the third one and that $p$-adic distance between
codons is a natural one in description of their information
content (the closer, the more similar).

\bigskip

\subsection{$p$-Adic genomic and bioinformation spaces}

Appropriateness of the $p$-adic codon space $\mathcal{C}_5 [64]$
to the genetic code is already shown in \cite{dragovich1a} and
will be reconsidered in the Section 3. Now we want to extend
$\mathcal{C}_5 [64]$ space approach to more general genetic and
bioinformation spaces.

Let us recall that four nucleotides are related to prime number
$5$ by their correspondence to the four nonzero digits $(1, 2, 3,
4)$ of $p = 5$. It is unappropriate to use the digit $0$ for a
nucleotide because it leads to non-uniqueness  in representation
of the codons by natural numbers. For example,  $123 = 123000$ as
numbers, but $123$ represents one and $123000$ two codons. This is
also a reason why we do not use $4$-adic representation for
codons, since it would contain a nucleotide presented by digit
$0$.  One can use $0$ as a digit to denote absence of any
nucleotide.

Let us note also that we have used on $\mathcal{C}_5 [64]$, in
\cite{dragovich1} and \cite{dragovich1a}, not only $5$-adic  but
$2$-adic distance as well.

\bigskip

{\bf Definition 1}. We shall call {\it $(p, q)$-adic genomic
space} a double $\Big( {\Gamma}_p\, \big[ (p - 1)^m \big]\,, \,
d_q \Big)$, where

\bea {\Gamma}_p\, \big[ (p - 1)^m \big]=\Big\{ n_0\, + n_1\, p \,
+ \cdots \, + \, n_{m-1}\, p^{m-1}\, :\nn \\ n_i = 1, 2, \cdots ,
p-1 \,, \,\, m\in \mathbb{N}\Big\}\, \label{2.5}\eea

\noi is the set of natural numbers, $d_q$ is the corresponding
$q$-adic distance on ${\Gamma}_p\, \big[ (p - 1)^m \big]$ and
nonzero digits $n_i$ are related to some $p - 1$ basic
constituents of a genomic system (or to any other biological
information system) in a unique way. Index $q$ is a prime number.

Here $m$ can be called also {\it multiplicity} of space elements
with respect to their constituents. In addition to $d_p$ there can
be a few other $d_q$ useful distances on ${\Gamma}_p\, \big[ (p -
1)^m \big]$.

For simplicity, we shall often call ${\Gamma}_p \,\big[ (p - 1)^m
\big]$ {\it $p$-adic genomic space} and  use notation

\beq n_0\, + n_1\, p \, + \cdots \, + \, n_{m-1}\, p^{m-1}
\,\equiv \, n_0\, n_1\,  \cdots \,  n_{m-1}\,, \label{2.6}\eeq

\noi where ordering of digits is in the opposite direction to the
standard one and seems here more natural. Earlier introduced codon
space $\mathcal{C}_5\, [64]$ can be regarded as a significant
example of the $p$-adic genomic spaces, i.e. $\mathcal{C}_5\, [64]
= \Gamma_5\, \big[ (5-1)^3 \big]$ as space of trinucleotides. Two
other examples, which will be used later, are: $\Gamma_5\, \big[ 4
\big]$ - space of nucleotides and $\Gamma_5\, \big[ 4^2 \big]$ -
space of dinucleotides.

Now we can introduce $p$-adic bioinformation space as a space
composed of some $p$-adic genomic spaces.

\bigskip

{\bf Definition 2}.  Let $\mathcal{B}_p\, [N]$ be a $p$-adic space
composed of $N$ natural numbers. We shall call $\mathcal{B}_p\,
[N]$ {\it $p$-adic bioinformation space} when it can be presented
as

\bea \mathcal{B}_p\, [N] \,\, \subset \prod_{m = m_1}^{m_2}
{\Gamma}_p \big[ (p - 1)^m \big]\,, \label{2.7}\eea

\noi where $m_1$ and $m_2$ are positive integers $(m_1 \leq m_2)$,
which determine the range of multiplicity  between $m_1$ and
$m_2$. In the sequel we shall present some concrete examples of
the $\mathcal{B}_p\, [N]$ spaces.

\bigskip

\subsection{DNA and RNA  spaces}

DNA sequences can be considered as a union of coding and
non-coding segments. Coding parts are composed of codons included
into genes, which are rather complex systems. In coding segments
is stored information, which through a series of complex processes
is translated into proteins. The space of coding DNA sequences
$(cDNA)$ can be presented as

\bea cDNA\, [N] \subset  \prod_{m=m_1}^{m_2}  \Gamma_{61} [60^m]
\,, \label{2.8}\eea

\noi where $p = 61$ because there are $60$ codons coding amino
acids (in the vertebral mitochondrial code). Thus $cDNA$ space can
be regarded as a set of $N$ coded sequences as well as a set of
$N$ discrete points (a lattice) of $\prod_{m=m_1}^{m_2}
\Gamma_{61} [60^m]$  space. While in $\mathcal{C}_5 \,[64]$ codons
are space elements, in $cDNA \, [N]$ they are building units.

The structure and function of non-coding sequences is still highly
unknown. They include information on various regulatory processes
in the cell. We assume that the space of non-coding DNA sequences
$(ncDNA)$ is a subspace

\bea ncDNA \subset  \prod_{m =m_1}^{m_2}  {\Gamma}_5 [4^m] \,,
\label{2.9}\eea

\noi where $m_1$ and $m_2$ are minimum and maximum values of the
size of non-coding segments.

In a similar way one can construct a space of all RNA sequences in
the cell.

\bigskip

\subsection{Protein space}

We mentioned  some basic properties of proteins in  Introduction.
Recall also that proteins functional properties depend on their
three-dimensional structure. There are four distinct levels of
protein structure (primary, secondary, tertiary and quaternary)
\cite{finkelshtein}. The primary structure is determined by the
amino acid sequence and the other ones depend on side chains of
amino acids (see Table \ref{Tab:01}). In addition to 20 standard
amino acids, presented in the Table \ref{Tab:01}, there are also 2
special nonstandard amino acids: selenocysteine and pyrrolysine
\cite{wiki}. They are also coded by codons, but are very rare in
proteins. Thus there are 22 amino acids encoded in the genetic
code. According to Jukes \cite{jukes} non-freezing code may
contain 28 amino acids.

The 20 standard (canonical) amino acids employed by the genetic
code  in proteins of the  living cells are listed in  Table
\ref{Tab:01}. Some their important chemical properties are
presented in  Table \ref{Tab:02}.

Now we want to construct an appropriate space whose elements are
proteins. We propose protein space $\mathcal{P}_p$ to be a
subspace of product of  genomic spaces

\bea
   \mathcal{P}_p\, [N] \subset  \prod_{m=m_1}^{m_2}  {\Gamma}_p \,
   [(p-1)^m]\,,
\label{2.10} \eea

\noi where the building units are amino acids. Thus $
\mathcal{P}_p\, [N]$ is a space of $N$ proteins with size measured
by the number of amino acids between $m_1$ and $m_2$ ($m_1 \sim
10$ and $m_2 \sim 10^4$).

In (\ref{2.10}) prime number $p$ is related to the number of amino
acids by relation: $p - 1 = $ {\it number of different amino
acids} used as building blocks in proteins. At present time there
are $22$ amino acids ($20$ standard and $2$ special) and
consequently $p = 23$ . One can argue that not all $22$ amino
acids have been from the very beginning of life and that has been
an evolution of amino acids. Namely, using 60 different criteria
for temporal order of appearance of the 20 standard amino acids
the  obtained result \cite{trifonov} is presented in Table
\ref{Tab:03}.  The first four amino acids (Gly, Ala, Asp and Val)
have the most production rate in Miller's experiment of  an
imitation of the atmosphere of the early Earth. This could
correspond  to $p = 5$ and single nucleotide codons in a primitive
code. In the case of dinucleotide code there are $16$ codons and
maximum amino acids that can be coded is 16, i.e. $p = 17 $. As we
already mentioned, according to Jukes \cite{jukes}, it is possible
to code 28 amino acids by trinucleotide code and it gives the
corresponding $p = 29$.

\bigskip

\section{$p$-Adic Genetic Code}

An intensive study of the connection between ordering of
nucleotides in  DNA (and RNA) and ordering of amino acids in
proteins led to the experimental deciphering of the standard
genetic code in the mid-1960s. The genetic code is understood as a
dictionary for translation of information from  DNA (through RNA)
to synthesis of proteins by amino acids. The information on amino
acids is contained in codons: each codon codes either an amino
acid or termination signal (see, e.g. Table \ref{Tab:03} as a
standard table of the vertebral mitochondrial code). To the
sequence of codons in  RNA corresponds quite definite sequence of
amino acids in a protein, and this sequence of amino acids
determines primary structure of the protein. The genetic code is
comma-free and non-overlapping. At the time of deciphering, it was
mainly believed that the standard code is unique, result of a
chance and fixed a long time ego. Crick \cite{crick} expressed
such belief in his "frozen accident" hypothesis, which has not
been supported by later observations. Moreover, it has been
discovered so far at least 16 different codes and found some
general regularities. At first glance the genetic code looks
rather arbitrary, but it is not. Namely, mutations between
synonymous codons give the same amino acid. When mutation alter an
amino acid then it is like substitution of the original by similar
one. In this respect the code is almost optimal.

\begin{table}
{{\bf TABLE 1.} \it List of 20 standard amino acids used in
proteins by living cells. 3-Letter and 1-letter abbreviations, and
chemical structure of their side chains are presented.}

\vskip3mm
 \label{Tab:01}
\centerline{ {\begin{tabular}{|l|l|l|}
 \hline \ & \ & \\
AMINO ACID & ABBR. & SIDE CHAIN (R)  \\
 \hline \ & \  & \\
 Alanine &  Ala, A  & -$CH_3$  \\
 Cysteine &  Cys, C  & -$CH_2$SH  \\
 Aspartate &  Asp, D  & -$CH_2$COOH   \\
 Glutamate &  Glu, E  & -$(CH_2)_2COOH$  \\
 Phenynalanine&  Phe, F  & -$CH_2C_6H_5$    \\
 Glycine &  Gly, G  & -$H$  \\
 Histidine &  His, H  & -$CH_2$-$C_3H_3N_2$   \\
 Isoleucine &  Ile, I  & -$CH(CH_3)CH_2CH_3$    \\
 Lysine &  Lys, K  & -$(CH_2)_4NH_2$  \\
 Leucine &  Leu, L  & -$CH_2CH(CH_3)_2$  \\
 Methionine & Met, M  & -$(CH_2)_2SCH_3$ \\
 Asparagine & Asn, N  & -$CH_2CONH_2$  \\
 Proline &  Pro, P  & -$(CH_2)_3$-  \\
 Glutamine &  Gln, Q  & -$(CH_2)_2CONH_2$   \\
 Arginine &  Arg, R  & -$(CH_2)_3NHC(NH)NH_2$  \\
 Serine &  Ser, S  & -$CH_2OH$  \\
 Threonine & Thr, T  & -$CH(OH)CH_3$  \\
 Valine &  Val, V  & -$CH(CH_3)_2$  \\
 Tryptophan &  Trp, W  & -$CH_2C_8H_6N$  \\
 Tyrosine &  Tyr, Y   & -$CH_2$-$C_6H_4OH$  \\
 \hline
\end{tabular}}{}}
\end{table}
%\end{center}
\bigskip

\bigskip

\begin{table}
{{\bf TABLE 2.} {\it Some chemical properties of 20 standard amino
acids. p + n is number of nucleons. These and some other recent
chemical values can be found in }\cite{wikipedia}.

\vskip3mm \label{Tab:02}} \centerline
{{\begin{tabular}{|l|r|l|l|c|}
 \hline \  & \\
  Amino & p+n & Polar & Hydro- & In proteins \\
  acids & {} & {} & phobic & \% \\
 \hline \   & \\
   Ala, \, A  & 89  & no & yes & 7.8  \\
   Cys, \, C  & 121 & no  & yes &  1.9 \\
   Asp, \, D  & 133 & yes & no   & 5.3 \\
   Glu, \, E  & 147 & yes & no   & 6.3 \\
   Phe, \, F  & 165 & no &  yes  & 3.9\\
   Gly, \, G  & 75  & no &  yes  & 7.2 \\
   His, \, H  & 155 & yes & no  &  2.3\\
   Ile, \,\, I  & 131 & no & yes  &  5.3 \\
   Lys, \, K  & 146 & yes& no   &  5.9\\
   Leu, \, L  & 131 & no & yes  &  9.1 \\
   Met, \, M  & 149 & no & yes  & 2.3\\
   Asn, \, N  & 132 & yes & no &  4.3\\
   Pro, \, P  & 115 & no & yes &  5.2 \\
   Gln, \, Q  & 146 & yes & no &  4.2 \\
   Arg, \, R  & 174 & yes & no &  5.1\\
   Ser, \, S  & 105 & yes & no  & 6.8\\
   Thr, \, T  & 119 & yes & no  & 5.9 \\
   Val, \, V  & 117 & no & yes &  6.6\\
   Trp, \, W  & 204 & no & yes &  1.4 \\
   Tyr, \, Y  & 181 & yes & yes & 3.2 \\
\hline
\end{tabular}}{}}
\end{table}
%\end{center}

\bigskip

\begin{table}
{{\bf TABLE 3.} { \it Temporal appearance of the 20 standard amino
acids} \cite{trifonov}.

\vskip3mm \label{Tab:03}} {\centerline{\begin{tabular}{llll}
\hline
(1) Gly & (2) Ala & (3) Asp & (4) Val\\
(5) Pro & (6) Ser & (7) Glu & (8) Leu\\
(9) Thr & (10) Arg & (11) Ile & (12) Gln \\
(13) Asn & (14) His & (15) Lys & (16) Cys\\
(17) Phe & (18) Tyr & (19) Met & (20) Trp\\ \hline
\end{tabular}}{}}
\end{table}

%\bigskip

Despite of  remarkable experimental successes, there is  no simple
and generally accepted theoretical understanding of the genetic
code. There are many papers in this direction (in addition to
already cited, see also, e.g. \cite{rumer} and \cite{swanson}),
scattered in various journals, with theoretical approaches based
more or less on chemical, biological and mathematical aspects of
the genetic code. Even before deciphering of the code there have
been very attractive theoretical inventions (of Gamow and Crick),
but the genetic code occurred to be quite different (for a review
on the early inventions around the genetic code, see
\cite{hayes}). However, the foundation of biological coding is
still an open problem. In particular, it is not clear why genetic
code exists just in  few known ways and not in many other possible
ones. What is a principle (or principles) employed in
establishment of a basic (mitochondrial) code? What are properties
of codons connecting them into definite multiplets which code the
same amino acid or termination signal? Answers to these and some
other questions should  lead us to discover an appropriate
theoretical model of the genetic code.

\bigskip

\begin{table}
{{\bf TABLE 4.}  \it The standard (Watson-Crick) table of the
vertebral mitochondrial code. Ter denotes the terminal (stop)
signal.

\vskip3mm \label{Tab:04}} \centerline{{\begin{tabular}{|l|l|l|l|}
 \hline \ & \ & \ & \\
  UUU \, Phe &   UCU \, Ser &  UAU \, Tyr &  UGU \, Cys  \\
  UUC \, Phe &   UCC \, Ser &  UAC \, Tyr &  UGC \, Cys  \\
  UUA \, Leu &   UCA \, Ser &  UAA \, Ter &  UGA \, Trp  \\
  UUG \, Leu &   UCG \, Ser &  UAG \, Ter &  UGG \, Trp  \\
 \hline \  & \  &  \ & \ \\
  CUU \, Leu &   CCU \, Pro &  CAU \, His &  CGU \, Arg   \\
  CUC \, Leu &   CCC \, Pro &  CAC \, His &  CGC \, Arg   \\
  CUA \, Leu &   CCA \, Pro &  CAA \, Gln &  CGA \, Arg   \\
  CUG \, Leu &   CCG \, Pro &  CAG \, Gln &  CGG \, Arg   \\
 \hline \  & \  & \  &   \\
  AUU \, Ile &   ACU \, Thr &  AAU \, Asn &  AGU \, Ser  \\
  AUC \, Ile &   ACC \, Thr &  AAC \, Asn &  AGC \, Ser  \\
  AUA \, Met &   ACA \, Thr &  AAA \, Lys &  AGA \, Ter  \\
  AUG \, Met &   ACG \, Thr &  AAG \, Lys &  AGG \, Ter  \\
 \hline \ & \   & \  &   \\
  GUU \, Val &   GCU \, Ala  &  GAU \, Asp &  GGU \, Gly  \\
  GUC \, Val &   GCC \, Ala  &  GAC \, Asp &  GGC \, Gly  \\
  GUA \, Val &   GCA \, Ala  &  GAA \, Glu &  GGA \, Gly  \\
  GUG \, Val &   GCG \, Ala  &  GAG \, Glu &  GGG \, Gly  \\
\hline
\end{tabular}}{}}
\end{table}

\bigskip

\begin{table}
{{\bf TABLE 5.}  \it Our table of the vertebral mitochondrial code
in the usual notation.

\vskip3mm \label{Tab:05}} \centerline{
 {\begin{tabular}{|l|l|l|l|}
 \hline \ & \ & \ & \\
  CCC \, Pro &   ACC \, Thr  &  UCC \, Ser &  GCC \, Ala  \\
  CCA \, Pro &   ACA \, Thr  &  UCA \, Ser &  GCA \, Ala  \\
  CCU \, Pro &   ACU \, Thr  &  UCU \, Ser &  GCU \, Ala  \\
  CCG \, Pro &   ACG \, Thr  &  UCG \, Ser &  GCG \, Ala  \\
 \hline \  & \  &  \ & \ \\
  CAC \, His &   AAC \, Asn  &  UAC \, Tyr &  GAC \, Asp  \\
  CAA \, Gln &   AAA \, Lys  &  UAA \, Ter &  GAA \, Glu  \\
  CAU \, His &   AAU \, Asn  &  UAU \, Tyr &  GAU \, Asp  \\
  CAG \, Gln &   AAG \, Lys  &  UAG \, Ter &  GAG \, Glu  \\
 \hline \  & \  & \  &   \\
  CUC \, Leu &   AUC \, Ile  &  UUC \, Phe &  GUC \, Val \\
  CUA \, Leu &   AUA \, Met  &  UUA \, Leu &  GUA \, Val \\
  CUU \, Leu &   AUU \, Ile  &  UUU \, Phe &  GUU \, Val \\
  CUG \, Leu &   AUG \, Met  &  UUG \, Leu &  GUG \, Val \\
 \hline \ & \   & \  &   \\
  CGC \, Arg &   AGC \, Ser  &  UGC \, Cys &  GGC \, Gly  \\
  CGA \, Arg &   AGA \, Ter  &  UGA \, Trp &  GGA \, Gly  \\
  CGU \, Arg &   AGU \, Ser  &  UGU \, Cys &  GGU \, Gly  \\
  CGG \, Arg &   AGG \, Ter  &  UGG \, Trp &  GGG \, Gly  \\
\hline
\end{tabular}}{}}
\end{table}
%\end{center}

\begin{table}
{{\bf TABLE 6.} \it Our $5$-adic table of the vertebral
mitochondrial code, which is a representation of the
$\mathcal{C}_5 \,[64]$ codon space.

\vskip3mm \label{Tab:06}}\centerline{{\begin{tabular}{|l|l|l|l|}
 \hline \ & \ & \ & \\
 111 \, Pro &  211 \, Thr  & 311 \, Ser & 411 \, Ala  \\
 112 \, Pro &  212 \, Thr  & 312 \, Ser & 412 \, Ala  \\
 113 \, Pro &  213 \, Thr  & 313 \, Ser & 413 \, Ala  \\
 114 \, Pro &  214 \, Thr  & 314 \, Ser & 414 \, Ala  \\
 \hline \  & \  &  \ & \ \\
 121 \, His &  221 \, Asn  & 321 \, Tyr & 421 \, Asp  \\
 122 \, Gln &  222 \, Lys  & 322 \, Ter & 422 \, Glu  \\
 123 \, His &  223 \, Asn  & 323 \, Tyr & 423 \, Asp  \\
 124 \, Gln &  224 \, Lys  & 324 \, Ter & 424 \, Glu  \\
 \hline \  & \  & \  &   \\
 131 \, Leu &  231 \, Ile  & 331 \, Phe & 431 \, Val \\
 132 \, Leu &  232 \, Met  & 332 \, Leu & 432 \, Val \\
 133 \, Leu &  233 \, Ile  & 333 \, Phe & 433 \, Val \\
 134 \, Leu &  234 \, Met  & 334 \, Leu & 434 \, Val \\
 \hline \ & \   & \  &   \\
 141 \, Arg &  241 \, Ser  & 341 \, Cys & 441 \, Gly  \\
 142 \, Arg &  242 \, Ter  & 342 \, Trp & 442 \, Gly  \\
 143 \, Arg &  243 \, Ser  & 343 \, Cys & 443 \, Gly  \\
 144 \, Arg &  244 \, Ter  & 344 \, Trp & 444 \, Gly  \\
\hline
\end{tabular}}{}}
\end{table}
%\end{center}

\bigskip

Let us now turn to   Table \ref{Tab:04}. We observe that this
table can be regarded as a big rectangle divided into 16 equal
smaller rectangles: 8 of them are quadruplets which one-to-one
correspond to 8 amino acids, and other 8 rectangles are divided
into 16 doublets coding 14 amino acids and termination (stop)
signal (by two doublets at different places). However there is no
 manifest symmetry in distribution of these quadruplets and
doublets.

In order to get a symmetry we have rewritten this standard table
into new one  by rearranging 16 rectangles. As a result we
obtained  Table \ref{Tab:05} which exhibits a symmetry with
respect to the distribution of codon quadruplets and codon
doublets. Namely, in our table quadruplets and doublets form
separately two figures, which are symmetric with respect to the
mid vertical line (a left-right symmetry), i.e. they are invariant
under interchange $C \leftrightarrow G$ and $A \leftrightarrow U$
at the first position in codons at all horizontal lines. Recall
that also  DNA is  symmetric under simultaneous interchange of
complementary nucleotides $C \leftrightarrow G$ and $A
\leftrightarrow T$ between its strands. All doublets in this table
form a nice figure which looks like letter $\mathbb{T}$.

Table \ref{Tab:06} contains the same distribution of amino acids
as Table \ref{Tab:05}, but codons are now presented by $5$-adic
numbers $n_0 n_1 n_2$ instead of capital letters (recall: C = 1,
A= 2,  U = 3, G = 4). This new table can be also regarded as a
representation of the $\mathcal{C}_5\, [64]$ codon space with
gradual increasing of integers from $1 1 1$ to $4 4 4$. The
observed left-right symmetry is now invariance under the
corresponding transformations $1 \leftrightarrow 4$ and $2
\leftrightarrow 3$. In other words, at each horizontal line one
can perform {\it doublet} $\leftrightarrow$ {\it doublet} and {\it
quadruplet} $\leftrightarrow$ {\it quadruplet} interchange around
vertical midline.

It is worth noting that  the above invariance leaves also
unchanged polarity and hydrophobicity  of the corresponding amino
acids in all but three cases: Asn $ \leftrightarrow $ Tyr, Arg $
\leftrightarrow $ Gly, and Ser $ \leftrightarrow $ Cys.

\bigskip

\subsection{Degeneracy of the genetic code}

Let us now explore distances between codons and their role in
formation of the genetic code degeneration.

To this end let us again  turn to  Table \ref{Tab:06} as a
representation of the $\mathcal{C}_5\, [64]$ codon space. Namely,
we observe that there are 16 quadruplets such that each of them
has the same first two digits. Hence $5$-adic distance between any
two different codons within a quadruplet is

\bea d_5 (a,\, b) = |a_0 + a_1 \, 5 + a_2 \, 5^2 - a_0 - a_1 \, 5
- b_2 \, 5^2 |_5 \nn
\\= |(a_2 - b_2) \, 5^2|_5 = |(a_2 - b_2)|_5 \,\, | 5^2 |_5 =
5^{-2}\,, \label{2.11} \eea because $a_0 = b_0$, $a_1 = b_1$ and
$|a_2 - b_2|_5 = 1$. According to (\ref{2.11}) nucleotides within
every quadruplet are at the smallest distance, i.e. they are
closest comparing to all other nucleotides.

Since codons are composed of three arranged  nucleotides, each of
which is either a purine or a pyrimidine, it is natural to try to
quantify similarity inside purines and pyrimidines, as well as
distinction between elements from these two groups of nucleotides.
Fortunately there is a tool, which is again related to the
$p$-adics, and now it is $2$-adic distance. One can easily see
that  $2$-adic distance between pyrimidines  C and U is $d_2 (1,
3) = |3 - 1|_2 = 1/2$ as the distance between purines  A and G,
namely $d_2 (2, 4) = |4 - 2|_2 = 1/2$. However $2$-adic distance
between C and A or G as well as distance between U and A or G is
$1$ (i.e. maximum).

  With respect to  $2$-adic distance, the above quadruplets may be regarded
 as composed of two doublets: $a = a_0\, a_1\, 1$ and $b = a_0\, a_1\, 3$
 make the first doublet, and
 $c = a_0\, a_1\, 2$ and $d = a_0\, a_1\, 4$ form the second one. $2$-Adic
 distance between codons within each of these doublets is
 $\frac{1}{2}$, i.e.
 \begin{equation}
d_2 (a,\, b) = |(3 -1)\, 5^2|_2 =\frac{1}{2} \,, \, \quad \, d_2
(c,\, d) = |(4 -2)\, 5^2|_2 =\frac{1}{2} \,,   \label{2.12}
 \end{equation}
because $3-1 = 4 - 2 = 2$.

One can now look at Table \ref{Tab:06} as a system of 32 doublets.
Thus 64 codons are clustered by a very regular way into 32
doublets. Each of 21 subjects (20 amino acids and 1 termination
signal) is coded by one, two or three doublets. In fact, there are
two, six and twelve amino acids coded by three, two and one
doublet, respectively. Residual two doublets code termination
signal.

Note that 2 of 16 doublets code 2 amino acids (Ser and Leu) which
are already coded by 2 quadruplets, thus amino acids Serine and
Leucine are coded by 6 codons (3 doublets).

To have a more complete picture on the genetic code it is useful
to consider possible distances between codons of different
quadruplets as well as between different doublets. Also, we
introduce distance between quadruplets or between doublets,
especially when distances between their codons have the same
value. Thus $5$-adic distance between any two quadruplets  in the
same column is $1/5$, while such distance between  other
quadruplets is $1$. $5$-Adic distance between doublets coincides
with distance between quadruplets, and this distance is
$\frac{1}{5^2}$ when doublets are within the same quadruplet.

The $2$-adic distances between codons, doublets and  quadruplets
are more complex. There are three basic cases: \begin{itemize}
\item codons differ only in one digit, \item codons differ in two
digits, \item codons differ in all three digits.
\end{itemize}
In the first case, $2$-adic distance can be
$\frac{1}{2}$  or $1$ depending whether difference between digits
is $2$ or not, respectively.

Let us now look at $2$-adic distances between doublets coding
leucine and also between doublets coding serine. These are two
cases of amino acids coded by three doublets. One has the
following distances:
\begin{itemize}
\item $d_2 (332, 334) = d_2 (132, 134) = \frac{1}{2}$  for
leucine, \item $d_2 (311, 241) = d_2 (313, 243) = \frac{1}{2}$
for serine.
\end{itemize}

If we use usual distance  between codons, instead of $p$-adic one,
then we would observe that two synonymous codons are very far (at
least 25 units), and that those which are close code different
amino acids. Thus we conclude that not usual metric but
ultrametric is inherent to codons.

How degeneracy of the genetic code is connected with $p$-adic
distances between codons? The answer is in the following {\bf
$p$-adic degeneracy principle}: {\it Two codons have the same
meaning with respect to amino acids if they are at smallest
$5$-adic and $0.5 \,$ $2$-adic distance}. Here $p$-adic distance
plays a role of similarity: the closer, the more similar. Taking
into account all known codes (see the next subsection) there is a
slight  violation of this principle. Now it is worth noting that
in modern particle physics just broken of the fundamental gauge
symmetry gives its standard model. There is a sense to introduce a
new principle (let us call it {\bf reality principle}): {\it
Reality is realization of some broken fundamental principles}. It
seems that this principle is valid not only in physics but also in
all sciences. In this context modern genetic code is an
evolutionary broken the above $p$-adic degeneracy principle.

\bigskip

\subsection{Evolution of the genetic code}

The origin  and early evolution of the genetic code are among the
most interesting and important  investigations  related to the
origin and whole evolution of the life. However, since there are
no concrete facts from that early period, it gives rise to many
speculations. Nevertheless, one can  hope that some of the
hypotheses may be tested looking for their traces in the
contemporary genomes.

It seems natural to consider biological evolution as an adaptive
development of simpler living systems to more complex ones.
Namely, living organisms are open systems in permanent interaction
with environment. Thus the evolution can be modelled by a system
with given initial conditions and guided by some internal rules
taking into account environmental factors.

We are going now to conjecture on the evolution of the genetic
code using our p-adic approach to the genomic space, and assuming
that preceding  codes used simpler codons and  older amino acids.

Recall that $p$-adic genomic space $\Gamma_p \, \big[ (p-1)^m
\big]$ has two parameters: $p$ - related to $p-1$ building blocks,
and $m$ - multiplicity of the building blocks in space elements.

\begin{itemize}
\item Case $\Gamma_2 \, \big[ 1  \big]$ is a trivial one and
useless for a primitive code. \item Case $\Gamma_3 \, \big[  2^m
\big]$ with $m =1, 2, 3$ does not seem to be realistic.

\item Case $\Gamma_5 \, \big[ 4^m  \big]$ with $m = 1, 2, 3$
offers a possible pattern to consider evolution of the genetic
code. Namely, the codon space could evolve in the following way:
$\Gamma_5 \, \big[ 4  \big] \to \Gamma_5 \, \big[ 4^2  \big] \to
\Gamma_5 \, \big[ 4^3  \big] = \mathcal{C}_5\, [64] $ (see also
Table \ref{Tab:07}).
\end{itemize}

According to  Table \ref{Tab:03} this primary code, containing
codons in the single nucleotide form (C, A, U, G), encoded the
first four amino acids: Gly, Ala, Asp and Val. From the last
column of  Table \ref{Tab:06} we conclude that the connection
between digits and amino acids is: 1 = Ala, 2 = Asp, 3 = Val, 4 =
Gly. In the primary code these digits occupied the first position
in the $5$-adic expansion (Table \ref{Tab:07}), and at the next
step, i.e. $\Gamma_5 \, \big[ 4  \big] \to \Gamma_5 \, \big[ 4^2
\big]$, they moved to the second position adding digits $1, 2, 3,
4$ in front of  each of them.

In $\Gamma_5 \, \big[ 4^2  \big]$ one has 16 dinucleotide codons
which can code up to 16 new amino acids. Addition of the digit $4$
in front of already existing codons $1, 2, 3, 4$ leaves their
meaning unchanged, i.e. 41 = Ala, 42 = Asp, 43 = Val,  and 44 =
Gly. Adding digits $3, 2, 1$ in front of the primary $1, 2, 3, 4$
codons one obtains 12  possibilities for coding some new amino
acids. To decide which amino acid was encoded by which of 12
dinucleotide codons, we use as a criterion  their immutability  in
the trinucleotide coding on the $\Gamma_5 \, \big[ 4^3  \big]$
space. This criterion  assumes that amino acids encoded earlier
are more fixed than those encoded later. According to this
criterion we decide in favor of the first row in each rectangle of
 Table \ref{Tab:06} and result is presented in  Table
\ref{Tab:08}.

Transition from  dinucleotide to trinucleotide codons occurred by
attaching nucleotides $1, 2, 3, 4$ at the third position, i. e.
behind each dinucleotide. By this way one obtains new codon space
$\Gamma_5 \, \big[ 4^3  \big] = \mathcal{C}_5\, [64]$, which is
significantly enlarged and  provides a pattern to generate known
genetic codes. This codon space $ \mathcal{C}_5\, [64]$ gives
possibility to realize at least three general properties of the
modern code: \vskip3mm

 (i) encoding of more than 16 amino acids,

(ii) diversity of codes,

(iii) stability of the gene expression.

\bigskip
Let us give some relevant clarifications.

(i) For functioning of contemporary living organisms it is
necessary to code at least 20 standard (Table \ref{Tab:01})  and 2
non-standard amino acids (selenocysteine and pyrrolysine).
Probably these 22 amino acids are also sufficient building units
for biosynthesis of all necessary contemporary proteins. While $
\Gamma_5 \, \big[ 4^2 \big]$ is insufficient, the genomic space $
\Gamma_5 \, \big[ 4^3 \big]$ offers approximately three codons per
one amino acid.

(ii) The eukariotic (often called standard or universal) code is
established around 1966 and was thought to be universal, i. e.,
common to all organisms. When the vertebral mitochondrial code was
discovered in 1979, it gave rise to believe that the code is not
frozen and that there are also some other codes which are mutually
different. According to later evidences, one can say that there
are at least 16 slightly different mitochondrial and nuclear codes
(for a review, see \cite{knight}, \cite{osawa} and references
therein). Different codes have some codons with  different
meaning. So, in the standard code there are the following changes
in Table \ref{Tab:06}: \begin{itemize}\item 232 (AUA): Met
$\rightarrow$ Ile, \item 242 (AGA) and 244 (AGG): Ter
$\rightarrow$ Arg, \item 342 (UGA): Trp $\rightarrow$ Ter.
\end{itemize}
Modifications in this 16 codes are not homogeneously distributed
on 16 rectangles of Table \ref{Tab:06}. For instance, in all 16
codes codons $4 1 i \,\,\, (i = 1, 2, 3, 4)$ have the same
meaning.

(iii) Each of the 16 codes is degenerate and degeneration provides
their stability against possible mutations. In other words,
degeneration helps to minimize codon errors.

Genetic codes based on single nucleotide and dinucleotide codons
were mainly directed to code amino acids with rather different
properties. This may be the reason why amino acids Glu and Gln are
not coded in dinucleotide code (Table \ref{Tab:08}), since they
are similar to Asp and Asn, respectively.   However, to become
almost optimal, trinucleotide codes have taken into account
structural and functional similarities of amino acids.

We presented here a hypothesis on the genetic code evolution
taking into account possible codon evolution, from 1-nucleotide to
3-nucleotide, and amino acids temporal appearance. This scenario
may be extended to the cell evolution, which probably should be
considered as a coevolution of all its main ingredients (for an
early idea of the coevolution, see \cite{wong}).

\bigskip

\begin{table}
{{\bf TABLE 7.}  \it $5$-Adic  system including digit $0$, and
containing single nucleotide, dinucleotide and trinucleotide
codons. If one ignores numbers which contain digit 0 in front of
any 1, 2, 3 or 4, then one has one-to-one correspondence between
1-digit, 2-digits, 3-digits numbers and single nucleotides,
dinucleotides, trinucleotides, respectively. It seems that
evolution of codons has followed transitions: single nucleotides
$\to$ dinucleotides $\to$ trinucleotides.

 \vskip3mm \label{Tab:07}}\centerline{ {\begin{tabular}{|l|l|l|l|l|}
 \hline \  & \   & \   & \ &  \\
 000   & 100 \bf C & 200 \bf A & 300 \bf U & 400 \bf G  \\
 \hline \  & \   &  \  & \ &  \\
 010 & 110 \bf CC & 210 \bf AC & 310 \bf UC & 410 \bf GC  \\
 020 & 120 \bf CA & 220 \bf AA & 320 \bf UA & 420 \bf GA  \\
 030 & 130 \bf CU & 230 \bf AU & 330 \bf UU & 430 \bf GU  \\
 040 & 140 \bf CG & 240 \bf AG & 340 \bf UG & 440 \bf GG \\
  \hline \  & \   &  \  & \ &  \\
 001 & 101 & 201 & 301 & 401  \\
  \hline \  & \   &  \  & \ &  \\
 011 & 111 \bf CCC & 211 \bf ACC & 311 \bf UCC & 411 \bf GCC  \\
 021 & 121 \bf CAC & 221 \bf AAC & 321 \bf UAC & 421 \bf GAC  \\
 031 & 131 \bf CUC & 231 \bf AUC & 331 \bf UUC & 431 \bf GUC  \\
 041 & 141 \bf CGC & 241 \bf AGC & 341 \bf UGC & 441 \bf GGC  \\
 \hline \  & \  &  \   & \ &  \\
 002 & 102 & 202 & 302 & 402  \\
  \hline \  & \   &  \  & \ &  \\
 012 & 112 \bf CCA & 212 \bf ACA & 312 \bf UCA & 412 \bf GCA  \\
 022 & 122 \bf CAA & 222 \bf AAA & 322 \bf UAA & 422 \bf GAA  \\
 032 & 132 \bf CUA & 232 \bf AUA & 332 \bf UUA & 432 \bf GUA  \\
 042 & 142 \bf CGA & 242 \bf AGA & 342 \bf UGA & 442 \bf GGA \\
 \hline \  & \   & \   & \ &  \\
 003 & 103 & 203 & 303 & 403  \\
  \hline \  & \   &  \  & \ &  \\
 013 & 113 \bf CCU & 213 \bf ACU & 313 \bf UCU & 413 \bf GCU  \\
 023 & 123 \bf CAU & 223 \bf AAU & 323 \bf UAU & 423 \bf GAU  \\
 033 & 133 \bf CUU & 233 \bf AUU & 333 \bf UUU & 433 \bf GUU  \\
 043 & 143 \bf CGU & 243 \bf AGU & 343 \bf UGU & 443 \bf GGU  \\
 \hline \  & \   & \   & \ &  \\
 004 & 104 & 204 & 304 & 404  \\
  \hline \  & \   &  \  & \ &  \\
 014 & 114 \bf CCG & 214 \bf ACG & 314 \bf UCG & 414 \bf GCG  \\
 024 & 124 \bf CAG & 224 \bf AAG & 324 \bf UAG & 424 \bf GAG  \\
 034 & 134 \bf CUG & 234 \bf AUG & 334 \bf UUG & 434 \bf GUG  \\
 044 & 144 \bf CGG & 244 \bf AGG & 344 \bf UGG & 444 \bf GGG  \\
\hline
\end{tabular}}{}}
\end{table}

\bigskip

\begin{table}
{{\bf TABLE 8.} \it The dinucleotide genetic code based on the
$p$-adic genomic space $\Gamma_5 \, [ 4^2 ]$. Note that it encodes
15 amino acids without stop codon, but encoding serine
twice.

\vskip3mm \label{Tab:08}} \centerline{ {\begin{tabular}{|l|l|l|l|}
 \hline \ & \ & \ & \\
 11 \, Pro &  21 \, Thr  & 31 \, Ser & 41 \, Ala  \\
  \hline \  & \  &  \ & \ \\
 12 \, His &  22 \, Asn  & 32 \, Tyr & 42 \, Asp  \\
  \hline \  & \  & \  &   \\
 13 \, Leu &  23 \, Ile  & 33 \, Phe & 43 \, Val \\
  \hline \ & \   & \  &   \\
 14 \, Arg &  24 \, Ser  & 34 \, Cys & 44 \, Gly  \\
 \hline
\end{tabular}}{}}
\end{table}
%\end{center}

\bigskip

\section{Concluding Remarks}

There are two aspects of the genetic code related to:

 \vskip3mm
(i)  multiplicity of codons which code the same amino acid,

(ii) concrete assignment of codon multiplets to particular amino
acids.

\bigskip
The above presented $p$-adic approach gives quite satisfactory
description of the aspect (i). Ultrametric behavior of $p$-adic
distances between elements of  the $\mathcal{C}_5 \,[64]$ codon
space radically differs from the usual ones. Quadruplets and
doublets of codons  have natural explanation within $5$-adic and
$2$-adic nearness. Degeneracy of the genetic code in the form of
doublets, quadruplets and sextuplets is direct consequence of
$p$-adic ultrametricity between codons. $p$-Adic $\mathcal{C}_5\,
[64]$ codon space is our theoretical pattern to consider all
variants of the genetic code: some codes are direct representation
of $\mathcal{C}_5 \, [64]$ and the others are its slight
evolutional modifications.

(ii) Which amino acid corresponds to which multiplet of codons? An
answer to this question should be expected from connections
between physicochemical properties of  amino acids and anticodons.
Namely, enzyme aminoacyl-tRNA synthetase links specific tRNA
anticodon and related amino acid. Thus there is no direct
interaction between amino acids and codons, as it was believed for
some time in the past.

%%%%%%%%%%%%%%%%%%%%%%%%%%%%%%%%%%%%%%%%%%%%%%%%%%%%%%%%%%%%%%%%%%%%%%%%%%%%%%%%
%\begin{figure}%figure2
%\centerline{\includegraphics{fig02.eps}}
%\caption{Caption, caption.}\label{fig:02}
%\end{figure}

%\begin{itemize}
%\item for bulleted list, use itemize \item for bulleted list, use
%\end{itemize}

%\vfill\pagebreak
%%%%%%%%%%%%%%%%%%%%%%%%%%%%%%%%%%%%%%%%%%%%%%%%%%%%%%%%%%%%%%%%%%%%%%%%%%%%%%%%%%%
%
%     please remove the " % " symbol from \centerline{\includegraphics{fig01.eps}}
%     as it may ignore the figures.
%
%%%%%%%%%%%%%%%%%%%%%%%%%%%%%%%%%%%%%%%%%%%%%%%%%%%%%%%%%%%%%%%%%%%%%%%%%%%%%%%%%%%%
%\begin{enumerate}
%\item this is item, use enumerate
%\end{enumerate}
%%%%%%%%%%%%%%%%%%%%%%%%%%%%%%%%%%%%%%%%%%%%%%%%%%%%%%%%%%%%%%%%%

Note that there are in general $4!$ ways to assign digits $1, 2,
3, 4$ to nucleotides C, A, U, G. After an analysis of all 24
possibilities, we have taken C = 1, \, A = 2,\, U = T = 3,\, G = 4
as a quite appropriate choice. In addition to various properties
already presented in this paper it exhibits also complementarity
of nucleotides in the DNA double helix  by relation C + G = A + T
= 5.

It would be useful to find an analogous connection between $22$
amino acids and digits $1, 2, \cdots , 22$ in $p = 23$
representation. Now there are $22!$ possibilities and to explore
all of them seems to be a hard task. However,  use of computer
analysis may help to find a satisfactory solution.

One can  express many above considerations of $p$-adic information
theory in linguistic terms and investigate possible linguistic
applications.

In this paper we have employed $p$-adic distances to measure
similarity between codons, which have been used to describe
degeneracy of the genetic code. It is worth noting that in other
contexts $p$-adic distances can be interpreted in quite different
meanings. For example, $3$-adic distance between cytosine and
guanine is $d_3 (1, 4) = \frac{1}{3}$, and between adenine and
thymine $d_3 (2, 3) =1$. This $3$-adic distance seems to be
natural to relate to hydrogen bonds between complements in DNA
double helix: the smaller distance, the stronger hydrogen bond.
Recall that  C-G and A-T are bonded by 3 and 2 hydrogen bonds,
respectively.

The translation of codon sequences into proteins is highly an
information-processing phenomenon. $p$-Adic information modelling
presented in this paper offers a new approach to systematic
investigation of ultrametric aspects of  DNA and RNA sequences,
the genetic code and the world of proteins. It can be embedded in
computer programs to explore $p$-adic side of the genome and
related subjects.

The above considerations and obtained results may be regarded as
contributions mainly towards foundations of (i) $p$-adic theory of
information and (ii) $p$-adic theory of the genetic code.

Summarizing,  contributions to

(i) $p$-adic theory of information contain:
\begin{itemize}
\item formulation of $p$-adic genomic space (whose examples are
spaces of nucleotides, dinucleotides and trinucleotides), \item
formulation of $p$-adic bioinformation space (whose examples are
DNA, RNA and protein spaces), \item relation between building
blocks  of information spaces and some prime numbers;
\end{itemize}

(ii) $p$-adic theory of the genetic code include:
\begin{itemize}
\item description of codon quadruplets and doublets by $5$-adic
and $2$-adic distances, \item observation of a symmetry between
quadruplets as well as between doublets at our table of codons,
\item formulation of degeneracy principle, \item formulation of
hypothesis on codon evolution.
\end{itemize}

Many problems remain to be explored in the future on the above
$p$-adic approach to genomics. Among the most attractive and
important themes  are: \begin{itemize}\item elaboration of the
$p$-adic theory of information towards genomics,  \item evolution
of the genome and the genetic code, \item structure and function
of non-coding DNA, \item ultrametric aspects of proteins, \item
creation of the corresponding computer programs.
\end{itemize}

\section*{Acknowledgement}
The work  on this paper was partially supported by the Ministry of
Science, Serbia, contract 144032D,  the Russian Foundation for
Basic Research, grant RFFI 06-04-08244, and Program of the Russian
Academy of Sciences "Dynamics of plants, animals and human
genofond". B.D. would like to thank M. Rakocevic for discussions
on chemical aspects of the genetic code.

\end{document}